# An Efficient Instance Segmentation Approach for Extracting Fission Gas Bubbles on U-10Zr Annular Fuel

Shoukun Sun[*,1], Fei Xu[*,2], Lu Cai[2], Daniele Salvato[2], Fidelma Dilemma[2], Luca Capriotti[2], Min Xian[#,1],Tiankai Yao[#,2]

[1] University of Idaho, Moscow, ID, United States

[2] Idaho National Laboratory, Idaho Falls, ID, United States

[*]Those authors contribute the same

[#]Corresponding authors

Min Xian: Email: mxian@uidaho.edu

Tiankai Yao, Email: Tiankai.yao@inl.gov

## Abstract

U-10Zr-based nuclear fuel is pursued as a primary candidate for next-generation sodium-cooled fast reactors. However, more advanced characterization and analysis are needed to form a fundamental understating of the fuel performance, and make U-10Zr fuel qualify for commercial use. The movement of lanthanides across the fuel section from the hot fuel center to the cool cladding surface is one of the key factors to affect fuel performance. In the advanced annular U-10Zr fuel, the lanthanides present as fission gas bubbles. Due to a lack of annotated data, existing literature utilized a multiple-threshold method to separate the bubbles and calculate bubble statistics on an annular fuel. However, the multiple-threshold method cannot achieve robust performance on images with different qualities and contrasts, and cannot distinguish different bubbles. This paper proposes a hybrid framework for efficient bubble segmentation. We develop a bubble annotation tool and generate the first fission gas bubble dataset with more than 3000 bubbles from 24 images. A multi-task deep learning network integrating U-Net and ResNet is designed to accomplish instance-level bubble segmentation. Combining the segmentation results and image processing step achieves the best recall ratio of more than 90% with very limited annotated data. Our model shows outstanding improvement by comparing the previously proposed thresholding method. The proposed method has promising to generate a more accurate quantitative analysis of fission gas bubbles on U-10Zr annular fuels. The results will contribute to identifying the bubbles with lanthanides and finally build the relationship between the thermal gradation and lanthanides movements of U-10Zr annular fuels. Mover, the deep learning model is applicable to other similar material micro-structure segmentation tasks.

# 1. Introduction

Next-generation advanced nuclear reactors with improved safety and economy are the future of nuclear energy for the United States and worldwide. The proposed fuel forms include U metal-based alloys, Tri-structural isotropic (TRISO) particles, and molten salts. The lack of a mechanistic understanding of irradiation behavior for these advanced fuels represents one of the leading reasons for the delay in their qualification. These complex systems may undergo microstructure transformation, phase redistribution, thermal property degradation in the fuel phase, and embrittlement, hardening and corrosion of the cladding and encapsulating materials during irradiation. The abovementioned phenomena are strongly interconnected and form a complex multi-factor problem. This problem, in turn, makes it difficult for conventional- empirical/physical-based fuel performance models to predict fuel behavior from estimated burnup and cladding temperature accurately. Taking the cladding strain as an example, the current BISON model revealed that the cladding radial dilation in HT9 was significantly less (80%) than the dilation measured during post-irradiation examination (PIE) [3].

U-10Zr-based metallic fuel is pursued as one of the leading candidate fuel forms for next-generation sodium-cooled fast reactors for the low fabrication cost and capability to achieve a higher burnup. Although thousands of U-10Zr fuel rods were irritated in the test reactors, such as Experimental Breeder Reactor II (EBR-II) and Fast Flux Test Facility (FFTF) from the 1960s to the 1990s [1,2], it has not been qualified for commercial use due to a lack of fundamental understanding of the nuclear fuel microstructure property evolution inside a reactor. One of the critical factors affecting fuel performance is the movement of lanthanides under a temperature gradient from the hot fuel center to the cool cladding surface. The lanthanide migration and its chemical interaction with cladding are critical for deteriorating the mechanical properties that could threaten fuel safety [4]. Moreover, Post-irradiation examinations (PIE) on advanced irradiated U-10Zr fuels in the Advanced Test Reactor of Idaho National Laboratory (INL) discovered that the lanthanide particles/nodules locate around the periphery of the pores [5-7]. Understanding the distribution changes of the pores in the cross-section of the advanced U-10Zr fuel will provide first-of-its-kind knowledge on the lanthanide transformation.

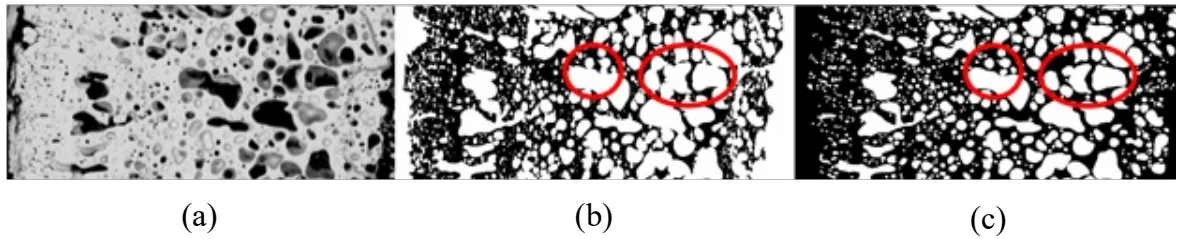

(a)             (b)             (c)

Figure 1. Sample results of fission gas bubble segmentation [11]. (a) A cross-section image patch of annular U-10Zr fuel; (b) segmentation results; and (c) annotated image.

Machine learning (ML) based models provide more reliable and accurate performance than traditional approaches, especially on tasks out of human capability. Many ML approaches can be found in literature exploring complex and large datasets to gain insights and accelerate scientific discoveries [8], such as accelerating testing to develop new materials [9], automated defect detection in Electron Microscopy [10], and so on. A new framework was proposed to segment and classify the fission gas bubbles in the (U, Zr) matrix regions of a U-10Zr advanced fuel [11]. In the framework, the authors applied image processing techniques to segment out the fission gas bubbles. A Decision-Tree model trained based on ~800 labeled bubbles was used to classify the segmentation bubbles into different categories. The techniques in the study can only segment out the bubbles on simple texture background (U, Zr) matrix regions without secondary UZr phase, and hard to get bubble boundaries, which will cause the misclassification, eventually obtaining inaccurate statistics of the number and size of bubbles as shown in **Figure 1**. Moreover, existing ML-based segmentation models achieved acceptable performance on natural images [12, 13], biomedical images [14], and material image [15], but detecting fission gas bubbles on the fuel cross-section is more challenging since the bubbles' color, size and shape vary greatly. Existing image processing techniques and pre-trained ML models cannot achieve good performance. Most ML models' performance heavily depends on the amount of annotated data used at the training stage. Advanced experimental characterization tools and modern imaging routinely provide high/ultrahigh-resolution images at an ever-increasing rate and volume. But it lacks sufficient and high-quality annotated training data.

In this paper, we proposed a hybrid framework for accurate and efficient fission gas bubble segmentation; and the contributions are summarized below.

1) The proposed hybrid segmentation decomposed the bubble segmentation task into two separated subtasks which only requires a small set of annotated bubbles during the training.

2) The proposed multitask instance segmentation network has a bubble region segmentation branch and boundary segmentation branch. It extracts and separates medium- and large-size bubbles accurately.

3) The proposed edge-based bubble segmentation approach generates accurate boundaries for small fission gas bubbles.

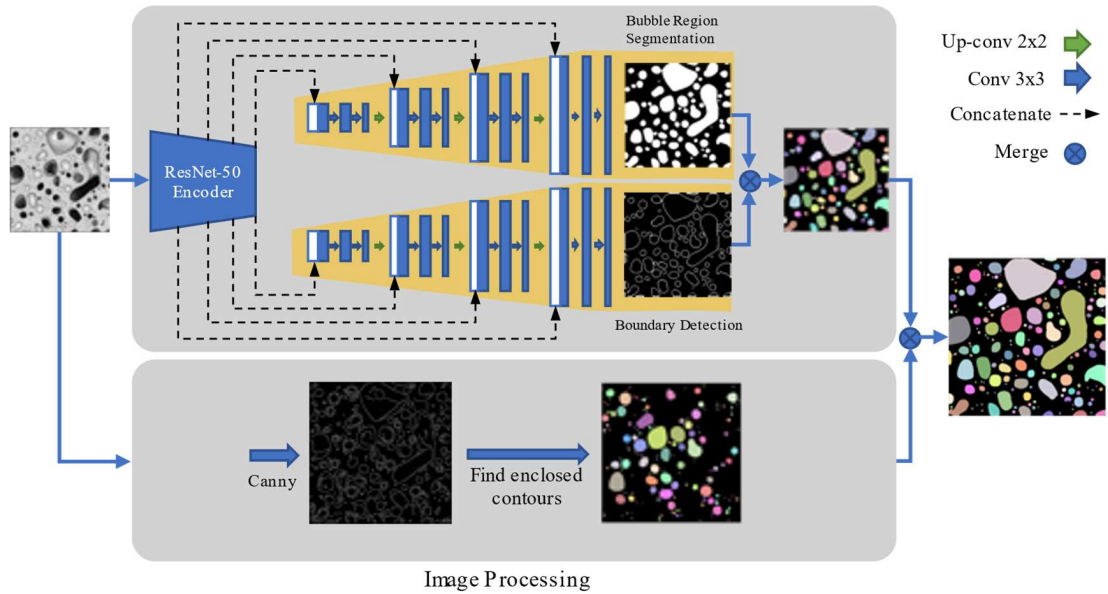

Figure 2. The proposed efficient instance segmentation framework.

## 2. Proposed Method

Three significant challenges exist when extracting bubbles from PIE images using deep learning-based approaches. First, it requires enormous time and effort to manually label a large dataset. In our PIE images, a great number of bubbles are unlabeled. Second, many tiny, black dot-liked bubbles widely exist in PIE images, and labeling the regions/boundaries of these bubbles is difficult. Third, existing instance segmentation approaches could be applied to extract and separate different bubbles. However, they, e.g., Mask-RCNN [22], are inefficient and inaccurate in segmenting large number of closely clustered objects. To address the challenges, we decompose the bubble segmentation tasks into two independent processing steps. In the first step, we propose a multitask instance segmentation network which is trained using a small annotated dataset to segment medium- and large-size bubbles. The second step is unsupervised and applies edge detection approach to extract small bubbles.

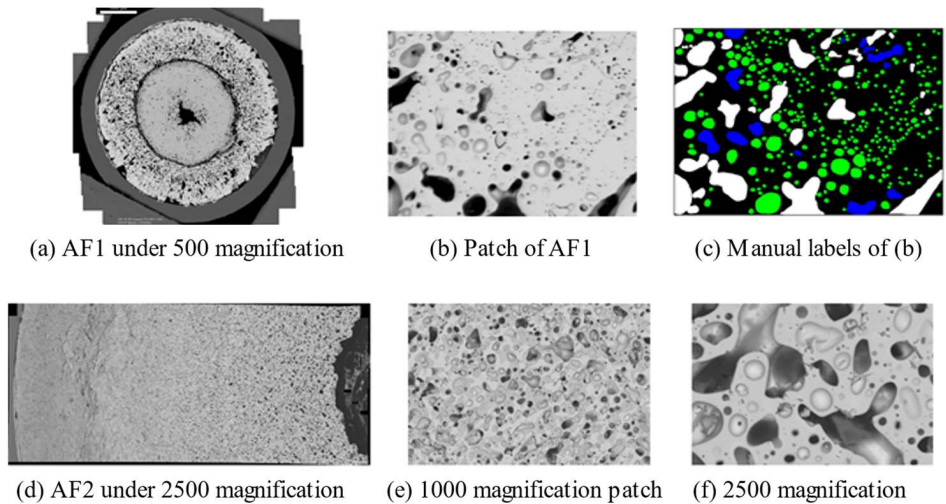

(a) AF1 under 500 magnification  (b) Patch of AF1  (c) Manual labels of (b)

(d) AF2 under 2500 magnification  (e) 1000 magnification patch  (f) 2500 magnification

Figure 3. Image samples from two advanced fuels under different magnifications.

## 2.1 Materials and data preparation

In recent years INL has been the leading national laboratory for research and development (R&D) on metallic fuel. Thanks to the development of advanced characterization capabilities at Material Fuel Complexity (MFC), including focused ion beam (FIB) sample preparation and probe-corrected transmission electron microscopy (TEM), local thermal conductivity microscopy (TCM), it is now possible to revisit the vast available PIE data we accumulated in the past and the newly established PIE data ranging from sub-nanometer to micrometer to obtain new findings. In a recent study, Cai et al. proposed a new framework on ~800 partially annotated bubbles on three 500-magnification image patches of a U-10Zr advanced fuel named AF [1], which are insufficient to obtain good performance for a DL model. In this study, we collected 585 Scanning electron microscope (SEM) image patches under 2500-magnification of a partial cross-section of another advanced U-10Zr fuel named AF2 [16]. The patches are collected from the hot center to the cladding, as shown in Fig. 3. To design a DL-based model, we need sufficient training data containing the original and annotated data. Moreover, the data of the two fuels under different magnifications will reveal the features of bubbles differently, for example, size, contour and texture, even for the same bubble. Under this circumstance, we developed an interactive annotation tool to label the fission gas bubbles. The annotated samples are shown in Figure 5.

**2.2 Multitask instance segmentation network for extracting medium- and large-size bubbles**

We propose a novel instance segmentation network (**Figure 2**) which treats each bubble as an instance and aims to extract and separate medium- and large-size bubbles from SEM images. As shown in Figure 2, the proposed network consists of one encoder and two decoder subnetworks. The encoder uses convolutional and pooling layers to extract meaningful features from input image at different scales. A ResNet-50 network is applied as the backbone network in the encoder. The first decoder is developed to segment bubble regions, and the second is to detect bubble boundaries. The results from the two decoders are combined to achieve an instance segmentation. The two decoders share the same feature input from the encoder. For preserving details, the intermedium feature maps of the encoder are passed to the corresponding layers in both decoders by skipping connections. These two decoders use the standard U-Net [22] decoder architecture.

The Dice loss function [21] is used in the bubble region segmentation branch. The Dice loss measures the quantitative difference between the region segmentation results ($\widehat{p_1}$) and the ground truth ($y_1$). It is defined by

$$L_{Dice}(y_1, \widehat{p_1}) = 1 - \frac{2y_1\widehat{p_1} + 1}{y_1 + \widehat{p_1} + 1}, \tag{1}$$

where $y_1$ is a 2D matrix that contains binary values in which value 1 denotes a bubble pixel, and value 0 represent a non-bubble pixel; and values in $\widehat{p_1}$ are the actual predictions produced by the regions segmentation network. The numerator and denominator are added to 1 as the smooth term to avoid division by zero.

The weighted binary cross-entropy loss function is used in the boundary detection branch. The loss function is given by

$$L_{WBCE}(y_2, \widehat{p_2}) = -[w_0(1-y_2)\log(1-\widehat{p_2}) + w_1 y_2 \log(\widehat{p_2})], \qquad (2)$$

where $y_2$ is a binary map that uses 1s to denote bubble boundary pixels; $\widehat{p_2}$ is the prediction of bubble boundaries; and $w_0$ and $w_1$ are the weights of boundary term and non-boundary terms, respectively. In experiments, $w_0$ is set to 0.1, and $w_1$ is set to 0.9.

The segmentation results are produced by subtracting bubble boundaries from bubble regions. The boundaries can disconnect touching bubble regions. During the post-processing, the final bubble instances are generated by connecting bubble pixels using the 8-adjacency system, the morphological dilation operation is applied to compensating the shrinking of bubble area.

### 2.3 Small bubble segmentation using edge detection

As shown in Figure 3, small bubbles are usually presented as black or grey dots in SEM images. These dots have homogeneous interior intensities. The grey bubbles have similar intensities to the background area; therefore, it is difficult to differentiate grey bubbles from background by using intensity thresholds. However, the grey bubbles have dark boundaries that clearly separate bubble regions and background, and it is more appropriate to use edge detection approaches to detect small bubbles.

In this work, we use the Canny edge detection [19] approach which includes five steps: 1) applying Gaussian filter; 2) calculating the intensity gradients; 3) applying non-maximum suppression to eliminate noise; 4) applying double thresholding to determine potential edges; and 5) suppressing weak edges. The Canny edge detection can high-quality edges and mitigate the impact of image noises.

The Canny edge detection is applied to extract small bubbles. Most medium- and large-size bubbles have long, irregular-shaped, and fuzzy boundaries; and edge detection approaches can only produce small, disconnected boundary pieces. But closed boundaries could be generated for small bubbles because they have more homogeneous boundary pixels. The final bubble regions are generated by applying the flood fill algorithm to fill the closed edges. Non-closed edges and large bubbles are removed from the final results.

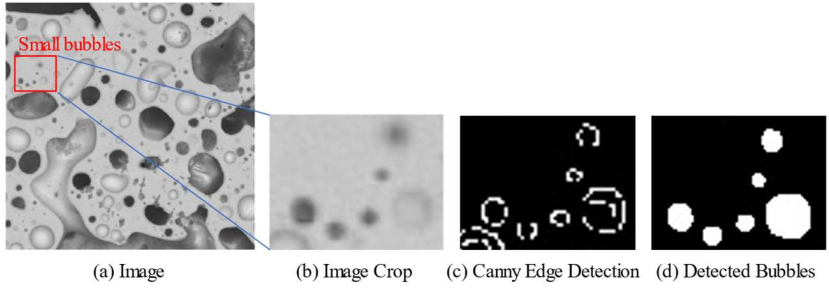

Figure 5. Samples of small bubble segmentation.

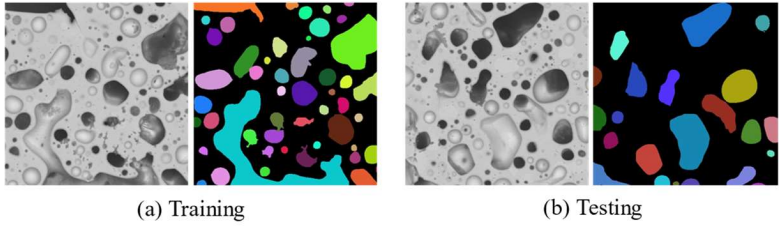

Figure 4. Image samples from the training and test sets.

## 3. Experimental Results

### 3.1 Setup

**Dataset**. The images are cropped into 515×512 non-overlapped patches. The training set contains 18 image patches and 827 bubbles, and only have ground truths for medium- and large-size bubbles. The test set contains 24 image patches and 685 bubbles.

**Training**. We adopt a ResNet-50 backbone as the encoder, and it is pretrained using ImageNet. In training, we use an AdamW [20] optimizer with a 0.001 learning rate and train the network for 100 epochs. An exponential learning rate scheduler with $\gamma = 0.97$ is used to decay the learning rate after every epoch. Training images are randomly augmented in every epoch with Gaussian blur, Gauss noise, brightness, contrast, horizontally/vertically flipping, scaling, and rotating. The batch size is set to 8.

**Evaluation metrics**. As the test image is partially labeled, we use the instance-level recall ratio $R_{iou}^I$ and the pixel-level ratio $R_{iou}^P$ to evaluate the performance. The recall ratio is defined by

$$R_{iou} = \frac{|TP|}{|P|}, \qquad (3)$$

where |TP| denotes the number of accurately segmented bubbles ($R^I_{iou}$) or pixels ($R^P_{iou}$) and |P| represents the number of total bubbles or pixels in ground truths. For the instance-level recall ratio $R^I_{iou}$, the total number of all labeled bubbles is treated as the |P|. Each ground truth bubble is paired with a predicted bubble with the largest IoU among all predictions. The |TP| counts all paired predictions that have IoU values with a ground truth greater than a threshold. A set of values ($[0.5, 0.6, 0.7, 0.8, 0.9]$) are used as thresholds in experiments. The pixel-level recall, $R^P_{iou}$, simply calculates the number of true bubble pixels over all labeled bubble pixels.

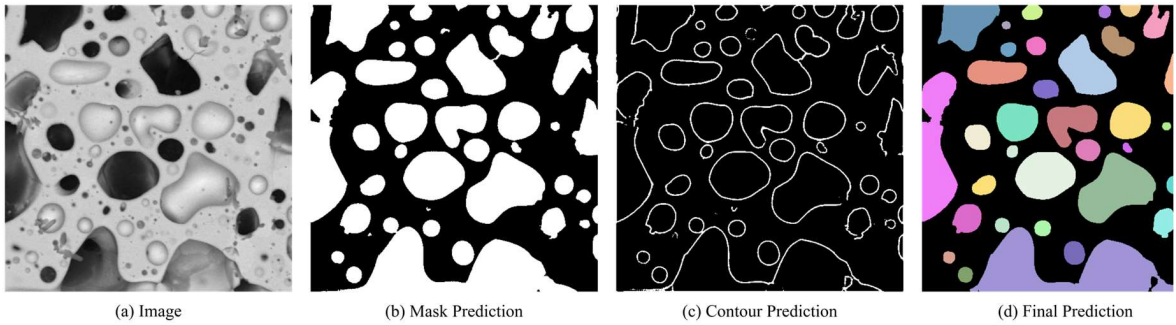

(a) Image     (b) Mask Prediction     (c) Contour Prediction     (d) Final Prediction

Figure 6. Sample results of the MTIS-Net.

### 3.2 MTIS-Network

Figure 6 shows an example of the outputs of the proposed MTIS-Net. The bubble region segmentation branch generate binary segmentation results for medium- and large-size bubbles. As shown in Figure 6(b), most bubbles are well segmented, but some bubbles are connected. Figure 7(c) shows the segmentation results of bubble boundaries generated by the boundary segmentation branch of MTIS-Net. The final segmentation results are shown in Figure 7(d), and different bubbles are illustrated using different colors.

### 3.3 Small bubble segmentation

Small bubbles are segmented using Canny edge detection approach. Figure 7(d) shows the results of the proposed small bubble segmentation approach; and most small bubbles have regular shapes. The merged results of the MTIS-Net and edge-based approach. Figure 7 (f) uses blue contours to show medium- and large-size bubbles identified by MTIS-Net, and red contours to demonstrate small bubbles detected by edge-based approach.

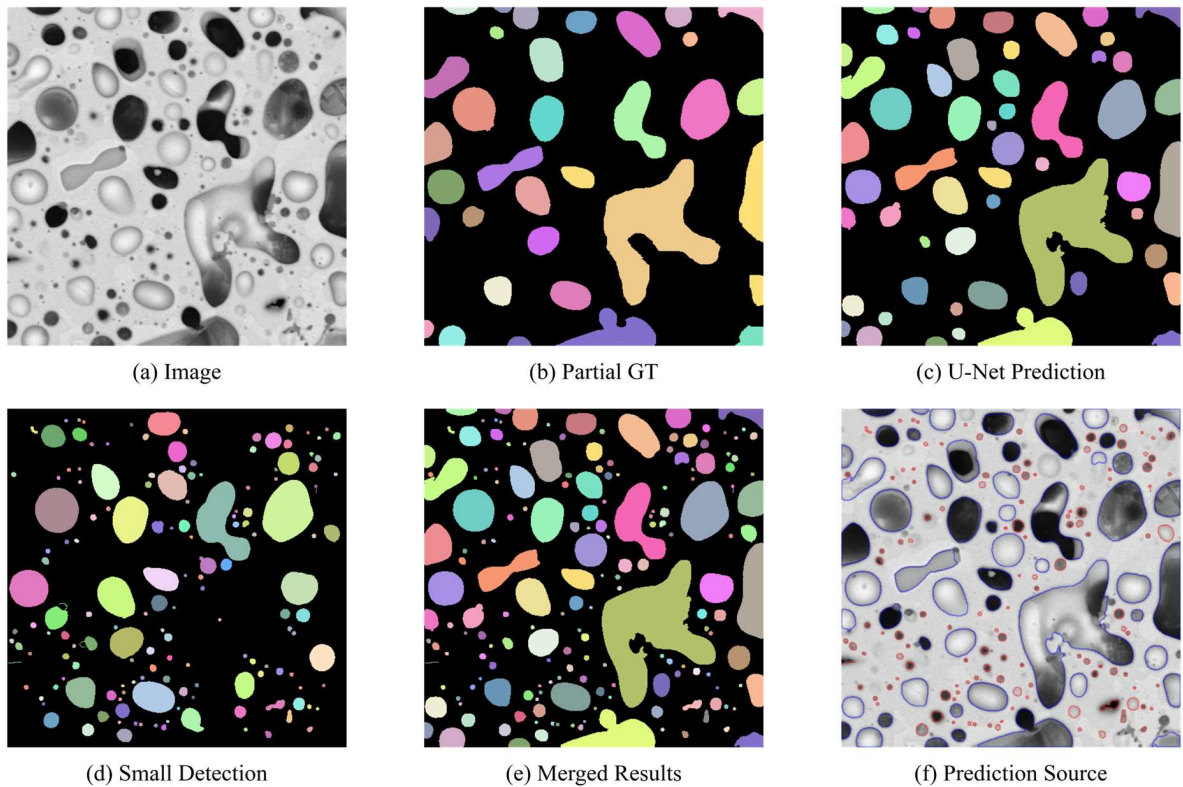

(a) Image  (b) Partial GT  (c) U-Net Prediction

(d) Small Detection  (e) Merged Results  (f) Prediction Source

Figure 7. Merged Results. (f) The blue contours are predicted by multi-task U-Net, and red are from small detection.

## 3.4 Overall performance

Table 1 demonstrates the instance-level recall ratios under different IoU thresholds (0.5, 0.6, 0.7, 0.8 and 0.9). In the "Bubble Numbers" table section, columns named "≥IoU threshold" show the number of bubbles that are successfully segmented based on specific IoU thresholds; and the "GT" column shows each image's total number of labeled objects. The "Instance-level Recall" section shows the corresponding recall values using different thresholds. In classic object localization and instant segmentation tasks, an IoU threshold of 0.5 is described as a loose criterion of the correct detection [17, 18]; and the value 0.75 is considered as "strict criterion"

[18].

Table 1 Instance-level Evaluation

| Img | Bubble Numbers | | | | | | Instance-level Recall | | | | |
|---|---|---|---|---|---|---|---|---|---|---|---|
| | ≥0.5 | ≥0.6 | ≥0.7 | ≥0.8 | ≥0.9 | GT | $R^I_{0.5}$ | $R^I_{0.6}$ | $R^I_{0.7}$ | $R^I_{0.8}$ | $R^I_{0.9}$ |
| 1 | 18 | 18 | 17 | 15 | 13 | 19 | 0.95 | 0.95 | 0.89 | 0.79 | 0.68 |
| 2 | 21 | 21 | 21 | 20 | 13 | 22 | 0.95 | 0.95 | 0.95 | 0.91 | 0.59 |
| 3 | 36 | 36 | 33 | 32 | 29 | 38 | 0.95 | 0.95 | 0.87 | 0.84 | 0.76 |
| 4 | 14 | 13 | 13 | 11 | 8 | 18 | 0.78 | 0.72 | 0.72 | 0.61 | 0.44 |
| 5 | 30 | 30 | 30 | 27 | 18 | 35 | 0.86 | 0.86 | 0.86 | 0.77 | 0.51 |
| 6 | 33 | 32 | 31 | 30 | 22 | 33 | 1.00 | 0.97 | 0.94 | 0.91 | 0.67 |
| 7 | 33 | 32 | 31 | 29 | 17 | 36 | 0.92 | 0.89 | 0.86 | 0.81 | 0.47 |
| 8 | 42 | 42 | 41 | 39 | 30 | 46 | 0.91 | 0.91 | 0.89 | 0.85 | 0.65 |
| 9 | 38 | 38 | 38 | 36 | 30 | 42 | 0.90 | 0.90 | 0.90 | 0.86 | 0.71 |
| 10 | 14 | 14 | 14 | 14 | 8 | 14 | 1.00 | 1.00 | 1.00 | 1.00 | 0.57 |
| 11 | 20 | 20 | 20 | 19 | 10 | 23 | 0.87 | 0.87 | 0.87 | 0.83 | 0.43 |
| 12 | 24 | 24 | 24 | 22 | 14 | 27 | 0.89 | 0.89 | 0.89 | 0.81 | 0.52 |
| 13 | 39 | 39 | 38 | 38 | 34 | 41 | 0.95 | 0.95 | 0.93 | 0.93 | 0.83 |
| 14 | 44 | 44 | 44 | 44 | 37 | 45 | 0.98 | 0.98 | 0.98 | 0.98 | 0.82 |
| 15 | 26 | 26 | 25 | 23 | 18 | 29 | 0.90 | 0.90 | 0.86 | 0.79 | 0.62 |
| 16 | 29 | 29 | 27 | 24 | 18 | 34 | 0.85 | 0.85 | 0.79 | 0.71 | 0.53 |
| 17 | 32 | 31 | 29 | 28 | 21 | 34 | 0.94 | 0.91 | 0.85 | 0.82 | 0.62 |
| 18 | 28 | 27 | 25 | 23 | 17 | 28 | 1.00 | 0.96 | 0.89 | 0.82 | 0.61 |
| 19 | 13 | 13 | 13 | 11 | 8 | 15 | 0.87 | 0.87 | 0.87 | 0.73 | 0.53 |
| 20 | 22 | 22 | 21 | 17 | 12 | 23 | 0.96 | 0.96 | 0.91 | 0.74 | 0.52 |
| 21 | 20 | 20 | 19 | 16 | 10 | 21 | 0.95 | 0.95 | 0.90 | 0.76 | 0.48 |
| 22 | 11 | 10 | 8 | 7 | 4 | 14 | 0.79 | 0.71 | 0.57 | 0.50 | 0.29 |
| 23 | 21 | 19 | 17 | 11 | 9 | 23 | 0.91 | 0.83 | 0.74 | 0.48 | 0.39 |
| 24 | 23 | 20 | 20 | 16 | 11 | 25 | 0.92 | 0.80 | 0.80 | 0.64 | 0.44 |
| **Total** | **631** | **620** | **599** | **552** | **411** | **685** | **0.92** | **0.91** | **0.87** | **0.81** | **0.60** |

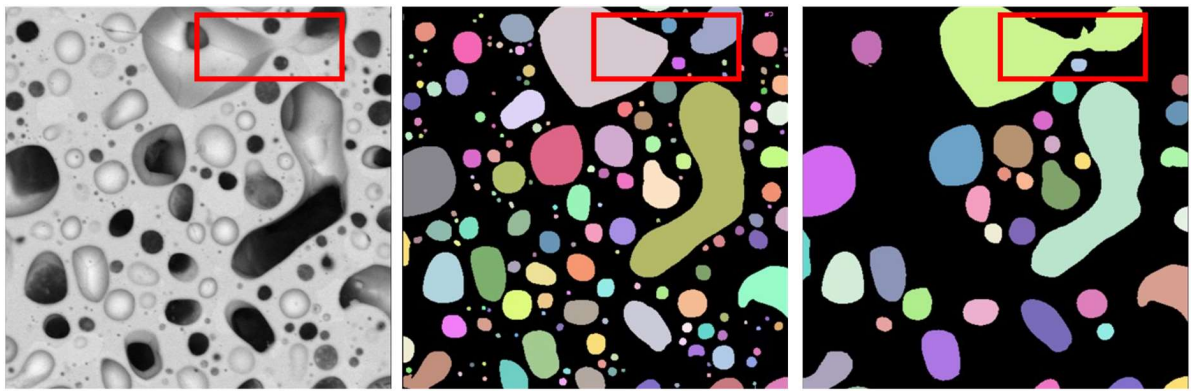

(a) Image     (b) Predicted     (c) Labeled

Figure 8. Difficulty of Defining a Bubble.

Table 2 Pixel-level recall ratio.

| Img | True Positive | Total Positive | Pixel-level Recall |
|---|---|---|---|
| 1 | 88344 | 91895 | 0.96 |
| 2 | 65683 | 68930 | 0.95 |
| 3 | 81535 | 86300 | 0.94 |
| 4 | 83962 | 99834 | 0.84 |
| 5 | 77886 | 81619 | 0.95 |
| 6 | 73899 | 78904 | 0.94 |
| 7 | 75386 | 81259 | 0.93 |
| 8 | 76543 | 82354 | 0.93 |
| 9 | 78035 | 82587 | 0.94 |
| 10 | 21374 | 23124 | 0.92 |
| 11 | 26084 | 28529 | 0.91 |
| 12 | 47586 | 52288 | 0.91 |
| 13 | 72939 | 76139 | 0.96 |
| 14 | 80045 | 83605 | 0.96 |
| 15 | 55762 | 59812 | 0.93 |
| 16 | 88650 | 95122 | 0.93 |
| 17 | 88710 | 93337 | 0.95 |
| 18 | 69115 | 73014 | 0.95 |
| 19 | 41549 | 44581 | 0.93 |
| 20 | 58103 | 64246 | 0.90 |
| 21 | 62136 | 68164 | 0.91 |
| 22 | 46198 | 51897 | 0.89 |
| 23 | 65970 | 74803 | 0.88 |
| 24 | 89642 | 97004 | 0.92 |
| **Total** | **1615136** | **1739347** | **0.93** |

Our method presents 0.92 recall with 0.5 IoU threshold and 0.60 recall with 0.90 IoU threshold. With the increasing IoU thresholds, the instance level recall ratios drop slowly, which shows the precision and stability of the proposed method.

Even though the label is created by experts, in specific situations, it is still hard to appropriately define if a bubble should be separated into two. Figure 8 shows such difficulty. The bubble inside the red box can be considered as one large bubble or two smaller bubbles. The expert labeled it a single bubble, but the model considered it two separate bubbles. Such a decision-making problem creates uncertainty in the instance-level recall. Hence, we also report the pixel-level recall to provide a more comprehensive evaluation.

Table 2 shows the pixel-level recall ratios for each image and the entire test set. The "True Positive" column is the number of pixels that are correctly classified. The "Total Positive" column represents each image's labeled bubble pixels. The total recall reaches 0.93. The highest and lowest recall of each image is 0.96 and 0.84, respectively.

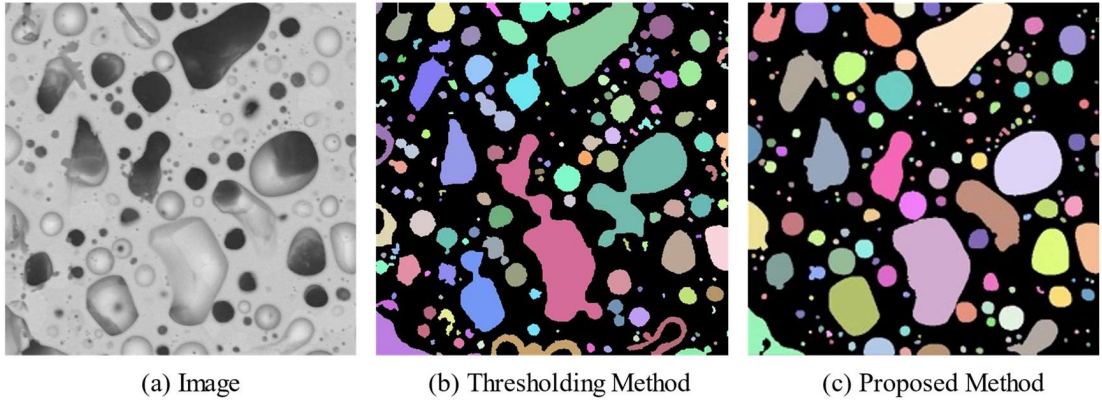

(a) Image  (b) Thresholding Method  (c) Proposed Method

Figure 9. Comparison of the thresholding and proposed methods.

Table 3. Instance-level Recall of Existing Method.

| Img | Bubble Numbers | | | | | | Instance-level Recall | | | | |
|---|---|---|---|---|---|---|---|---|---|---|---|
| | $\geq 0.5$ | $\geq 0.6$ | $\geq 0.7$ | $\geq 0.8$ | $\geq 0.9$ | GT | $R^I_{0.5}$ | $R^I_{0.6}$ | $R^I_{0.7}$ | $R^I_{0.8}$ | $R^I_{0.9}$ |
| Thresholding | 371 | 283 | 184 | 98 | 19 | 685 | 0.54 | 0.41 | 0.27 | 0.14 | 0.03 |
| **Ours** | **631** | **620** | **599** | **552** | **411** | **685** | **0.92** | **0.91** | **0.87** | **0.81** | **0.60** |

## 3.5 Comparison with existing work

In the previous study [11], the bubbles were segmented with a pure image processing process that utilized thresholding method. In this section, we compare the thresholding method and the proposed method on our dataset. As shown in Figure 9, the thresholding method tends to over-segment the bubbles, and the proposed method can generate more accurate results. Table 3 shows the instance-level recalls ratio of the thresholding method and proposed method. The recall ratio of the thresholding method reaches 0.54 with the loose criterion 0.5. Meanwhile, its recall is only 0.03 with the 0.9 IoU threshold, and the thresholding method has poor performance in segmenting objects precisely. Compared with the instance-level recalls in Table 1, the recalls at 0.5 IoU of the proposed method is 0.92. The improvement of the proposed method reaches 70%.

## 4. Discussions

The expensive cost of preparing ground truths is one of the major challenges in ML-based fission gas bubble segmentation, especially labeling tiny bubbles. The training images are not fully labeled. Part of

medium-, large-sized and all small bubble areas are marked as background during training. This circumstance hinders the model from learning the concept of the targeted object as the training data somewhat provides false knowledge.

Due to the incomplete ground truth, the model's performance is evaluated by recall ratio. We cannot conduct a comprehensive evaluation using more conventional instance or semantic segmentation metrics, such as mean average precision (mAP) and IoU. The drawback of recall is that it only counts on provided ground truths but cannot fully reveal the performance with the occurrence of over-segmentation.

The completely labeled ground truth with any sized bubbles is a solution to overcome the existing defects of training and evaluation. However, creating a large number of precise labels for training and evaluation is challenging. A more feasible way is to develop a semi-supervised model that can learn from incompletely labeled images. So that we can train a model on a partially labeled set and evaluate it on a smaller, fully labeled set.

## 5. Conclusion

In this study, we propose an instance-level PIE bubble segmentation approach. The proposed approach consists of an image processing step and a multi-task U-Net model for dealing with different-sized bubbles. The proposed method reaches good performance with very limited ground truths. Our model shows outstanding improvement by comparing the previously proposed thresholding method. The better performance provides more accurate fission gas bubble quantitative results which impact the distribution of fission gas bubbles' categories, especially the bubbles with lanthanide. The model will be unitized on the other U-10Zr annular fuels and contribute to build the relationship between thermal conductivity and lanthanide movements. Moreover, the proposed method has promising to apply to many material micro-structure segmentation tasks.